\documentclass{KapProc} 
\normallatexbib

\startauthorindex

\usepackage{graphicx}
\usepackage{dcolumn}
\usepackage{bm}
\usepackage{epsf}

\begin{document}

\articletitle{Ferroelectric Phase Transitions in a Lattice
Pseudo-Jahn-Teller Model}

\author{Dennis P.~Clougherty}
\affil{Department of Physics\\
University of Vermont\\
Burlington, VT 05403-0125}
\email{dpc@physics.uvm.edu}


\begin{abstract}
A lattice generalization of the $(b_{1u}\oplus b_{2g})\otimes b_{3u}$ pseudo-Jahn-Teller
model is analyzed within mean-field theory.  Ferroelectric or
antiferroelectric ground states are found for sufficiently
large pseudo-Jahn-Teller coupling.
\end{abstract}

\section*{Introduction}
Inspired by the family of fullerene-based polymeric structures synthesized
\cite{pekker,stephens}, a model
describing vibronic effects in a linear chain of molecules is introduced and analyzed
within mean-field theory.  The zero-temperature phase diagrams at carrier concentrations
$n=1$ and $2$ are presented for this model.  Only a local $(b_{1u}\oplus b_{2g})\otimes b_{3u}$
 pseudo-Jahn-Teller interaction is considered here; spin and charge fluctuation effects are
ignored.

The local symmetry of the molecule is taken to be D$_{2h}$ for simplicity;  the set of
$b$-representations transform as components of a vector.
The electronic structure of the molecule is taken to have a $b_{1u}$ level as the HOMO
 and a $b_{2g}$ level as the LUMO.    A distortion of $b_{3u}$ symmetry will
vibronically couple these levels.  For a molecule with some
ionic bonding character, such a $b_{3u}$ distortion produces an electric dipole
moment in proportion to the displacement\cite{clougherty}. It is the ordering of these local
dipole moments that is the concern of this work.

\section{The Hamiltonian}

The Hamiltonian is given by
\begin{equation}
{\cal H}={\cal H}_{0}+{\cal H}_{{\rm PJT}}+{\cal
H}_{{\rm elast}}
\label{ham}
\end{equation}
where
\begin{equation}
{\cal H}_{0}=-t\sum_{n}  (c^\dagger_{1,
n} c^{\phantom{\dagger}}_{1, n+1}+ c^\dagger_{2,
n} c^{\phantom{\dagger}}_{2, n+1}+{\rm h.c.})+
\sum_{n}  \Delta c^\dagger_{2, n} c^{\phantom{\dagger}}_{2, n}
\label{band}
\end{equation}
where $n$ runs over molecule sites, $c^{{\dagger}}_{1, n}$
($c^{\phantom{\dagger}}_{1, n}$) creates (annihilates) an electron
in orbital $b_{1u}$ at site $n$, $c^{{\dagger}}_{2, n}$
($c^{\phantom{\dagger}}_{2, n}$) creates (annihilates) an electron
in orbital $b_{2g}$ at site $n$, and $\Delta$ is the energy
spacing between the $b_{1u}$ and $b_{2g}$ orbitals. The hopping
matrix is taken to be diagonal in orbital type and spin, and the
spin indices are suppressed for clarity.

The pseudo-Jahn-Teller Hamiltonian on the $n$th molecular site is
\begin{equation}
{\cal H}_{{\rm PJT}}=  -{g} \sum_{n} (c^\dagger_{1, n}
c^{\phantom{\dagger}}_{2, n} Q_{n}+{\rm h.c.}) \label{pjt}
\end{equation}
where $g$ is the pseudo-Jahn-Teller coupling strength, and $Q_n$ is a
displacement of $b_{3u}$ symmetry on the $n$th
molecule.  The magnitude of the induced local electric dipole  is proportional  to $Q_n$. Hence,
a ferrodistortive state
is equated with a ferroelectric one.

${\cal H}_{{\rm elast}}$ is the elastic energy for the molecular
displacements.
\begin{equation}
{\cal H}_{{\rm elast}}={\kappa\over 2}\sum_n Q_{n}^2
\label{elastic}
\end{equation}

The Hamiltonian can be written in $k$-space as
\begin{equation}
{\cal H}= \sum_{\alpha, k}{\epsilon_\alpha}(k) c^\dagger_{\alpha,
k} c^{\phantom{\dagger}}_{\alpha, k} -{g\Delta\over\sqrt{N}}
\sum_{ k, q}  (c^\dagger_{1, k+q} c^{\phantom{\dagger}}_{2, k}
Q_{q}+{\rm h.c.})
+ {1\over 2}\kappa\sum_{q} |Q_{ q}|^2
\label{kham}
\end{equation}
where $\epsilon_{g}(k)=\Delta-2t\cos k$ and $\epsilon_{u}(k)=-2t\cos k$.

\section{Mean Field Theory}

With a single mode of distortion given by a non-vanishing $Q_{q}$,
${\cal H}$ can be diagonalized with a canonical transformation so that
\begin{equation}
{\cal H}=\sum_k (E_{k+} n_{k+}+E_{k-} n_{k-})+
{\kappa\over 2}|Q_{q}|^2
\end{equation}
where
\begin{equation}
E_{k\pm}=\frac{1}{2}(\epsilon_u(k)+\epsilon_g(k+q))\pm \frac{1}{2}
\sqrt{(\epsilon_g(k+q)-\epsilon_u(k))^2+{4g^2\over
N}|Q_q|^2}
\end{equation}
and $n_{k\pm}$ are the usual occupation number operators of the
vibronic quasiparticles.

The ground state energy of a chain with $n$ carriers ($0 \le n\le 2$) per molecule
on average is
obtained by fractionally filling the lowest energy vibronic band.
In the continuum limit,
\begin{equation}
E(Q_q)={N\over \pi}\int_{-{\pi n/2}}^{\phantom{-}\pi n/2} dk\
E_{k-} +{\kappa\over 2}|Q_{q}|^2
\end{equation}

Minimizing $E(Q_q)$ with respect to the distortion gives the mean-field
equation to determine
the magnitude of $Q_q$.
\begin{equation}
{\pi\kappa\over 2 g^2}=\int_{-{\pi n/2}}^{\phantom{-}\pi n/2} dk\
{1\over\sqrt{{4g^2\over N}|Q_q|^2+(\Delta+4t\sin{q/2}\sin(k+q/2))^2}}
\end{equation}
For sufficiently small coupling, this equation can not be satisfied.  There is a $q-$dependent
critical
coupling strength $g_c$ given by
\begin{equation}
g^2_c(q)={{\pi\kappa\Delta\over 2}\over\int_{-{\pi n/2}}^{\phantom{-}\pi n/2}
{ dk\over 1+u\sin q/2 \sin(k+q/2)}}
\end{equation}
where $u={4t\over\Delta}$.
For values of $g$ above $g_c(q)$, a mean-field distortion of wavenumber $q$ sets in;
for values
of $g$ below $g_c(q)$, there is no such distortion.

\section{The Phase Diagram}

\begin{figure}[ht]
\epsfbox[0 0 288 177]{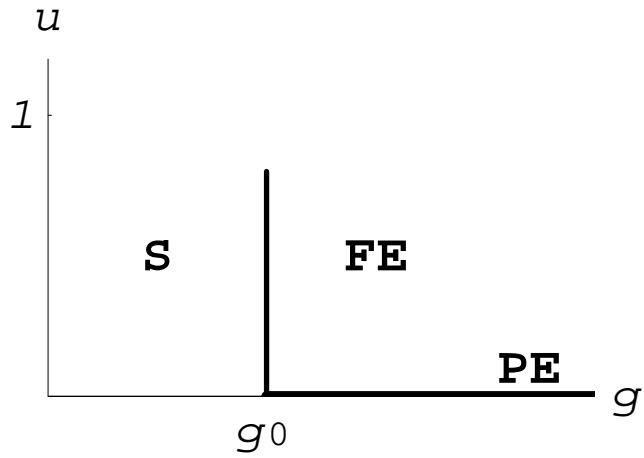}
\vspace{12pt}
\caption{Zero-temperature phase diagram for $n=1$.}
\label{phase}
\end{figure}

For the case of an average occupancy of one carrier per molecule, the ground state is found to
have a $q=0$ distortion for $g > g_0 (=\sqrt{\kappa\Delta\over 2})$, consequently breaking inversion
symmetry of the chain.  The phase diagram is pictured in Fig.~\ref{phase}.

\begin{figure}
\epsfbox[0 0 288 177]{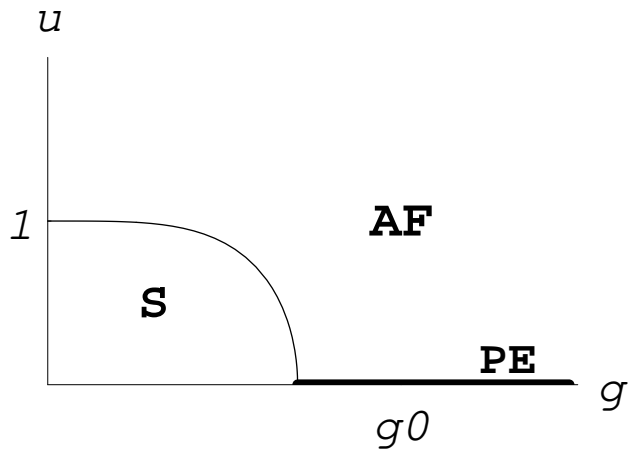}
\caption{Zero-temperature phase diagram for $n=2$.}
\label{afphase}
\end{figure}

For the case of two carriers per molecule, the ground state has a $q=\pi$ distortion when
\begin{equation}
g^2 > {g_0^2\over 2}\sqrt{1-u^2}
\end{equation}
The results are summarized in Fig.~\ref{afphase}.

\begin{acknowledgments}
This work was supported by the National Science Foundation through a
grant for the
Institute for Theoretical Atomic and Molecular Physics at Harvard
University and
the Smithsonian Astrophysical Observatory.
\end{acknowledgments}

\end{document}